\documentclass[prl,amsmath,amssymb,showpacs]{revtex4}
\usepackage{graphicx}

\begin{document}

\title{Supersymmetric Skyrmions : Numerical Solutions}

\author{Noriko Shiiki}
\email{norikoshiiki@mail.goo.ne.jp}
\author{Nobuyuki Sawado}
\email{sawado@ph.noda.tus.ac.jp}
\author{Shinsho Oryu}
\email{oryu@ph.noda.tus.ac.jp}

\affiliation{Department of Physics, Tokyo University of Science, Noda, Chiba 278-8510, Japan}

\date{\today}

\begin{abstract}
We consider the ${\cal N}=1$ Skyrme model and obtain 
supersymmetric skyrmion solutions numerically. 
The model necessarily contains higher derivative 
terms and as a result the field equation becomes a fourth-order differential 
equation. Solving the equation directly leads to runaway solutions as expected 
in higher derivative theories. 
We, therefore, apply the perturbation method and show that skyrmion solutions 
exist upto the second order in the coupling constant.   
\end{abstract}


\maketitle 
\section{1. Introduction}
Soliton solutions in non- or supersymmetric gauge theories at large-$N$ 
have been playing an important role to understand non-perturbative 
phenomena in QCD. 

A well-known example is the Skyrme model and its soliton solutions, 
called skyrmions~\cite{skyrme58}. Witten investigated mesons and glue states in the 
large-$N$ limit of QCD by a systematic expansion in powers of $1/N$ and showed that 
baryons emerge as solitons with mass of $O(N)$~\cite{witten79}. 
In the successive papers~\cite{witten83}, it was shown that the resultant 
effective theory is the Skyrme model and skyrmions are interpreted as 
baryons~\cite{adkins83}.

More recently, it was found that domain wall solutions exist in large-$N$ SUSY 
gluodynamics (SQCD)~\cite{dvali99} as well as in non-SUSY QCD~\cite{witten98,shifman99}. 
These walls behave as D-branes on which the string could end if 
they are BPS saturated~\cite{witten97}. An interesting observation is that the wall 
width is $O(1/N)$ and correspondingly heavy states are expected to emerge as solitons 
with mass of $O(N)$~\cite{gabadadze00}.  

In this context, it may be natural to consider that solitons 
with mass of $O(N)$ in SQCD could have something to do with skyrmions in the 
supersymmetric version of the Skyrme model.  
The supersymmetric Skyrme model was constructed in Ref.~\cite{bergshoeff85}. 
The extension to supersymmetry restricts us to work on a $CP(1)$ 
target space rather than $S^{3}$. The authors concluded that the possibility of 
the existence of soliton solutions is not excluded when the higher derivative 
terms are take into account. 

In this paper we construct supersymmetric skyrmion solutions numerically.  
In the supersymmetric case, the Skyrme field equation is not second-order but 
fourth-order in derivatives. In general, higher derivative theories lack a 
lowest-energy state and exhibit runaway solutions along with physical 
solutions, which makes numerical computation unstable. 
The prescription was proposed by Simon in Ref.~\cite{simon90}. 
According to his argument, if higher-order derivative terms can be considered as a 
small perturbation, physical solutions should be Taylor-expandable  
around the leading order solution.     
We shall apply this perturbation method to the supersymmetric Skyrme field 
equation and obtain soliton solutions upto the second order in the coupling constant. 
The dependence of the skyrmion solutions on the coupling constant is also examined. 

\section{2. The Supersymmetric Skyrme Model}
In this section, we give a brief review of the ${\cal N}=1$ supersymmetric Skyrme model 
constructed in Ref.~\cite{bergshoeff85}.
In supersymmetric theories, the target manifold ${\cal M}$ must be K{\"a}hlar~\cite{zumino79}.  
It was shown in Ref.~\cite{bergshoeff85} that the only nontrivial homotopy 
group in four dimensional spacetime is $\pi_{3}(CP(1))=Z$. 
The complex projective space $CP(N)$ is realised by gauging the $U(1)$ subgroup of 
$SU(N)$, {\it i.e.} $CP(N) \equiv SU(N)/SU(N-1)\times U(1)$. 
This amounts to replacing ordinary derivatives in the Lagrangian into 
covariant derivatives $D_{\mu}=\partial_{\mu}-iV_{\mu}U\tau_{3}$ where 
$\tau_{i}$ $(i=1,2,3)$ are Pauli matrices. The Skyrme Lagrangian 
is then given by 
\begin{eqnarray}
	{\cal L}=-\frac{f_{\pi}^{2}}{16}{\rm tr}(D^{\mu}U^{\dagger}
	D_{\mu}U)+\frac{1}{32e^{2}}{\rm tr}[U^{\dagger}D_{\mu}U, 
	U^{\dagger}D_{\nu}U]^{2} \,. \label{skyrme_action}
\end{eqnarray}
where $f_{\pi}$ is a pion decay constant and $e$ is a dimensionless 
constant. Let us parameterise the chiral field in terms of the 
complex scalars $A=(A^{1},A^{2}) \in C^{2}$ with 
${\bar A}A\equiv A_{1}^{*}A_{1}+A_{2}^{*}A_{2}=1$. 
$A_{i}$ is related to $SU(2)$ matrix by 
\begin{eqnarray}
U=\left(
\begin{matrix}
	A_{1} & -A_{2}^{*} \\
	A_{2} &  A_{1}^{*} \label{}
\end{matrix}
\right)\,.
\end{eqnarray}
One can parameterise the gauge field in terms of $A_{i}$ as     
\begin{eqnarray}
	V_{\mu}=-\frac{i}{2}[{\bar A}\partial_{\mu}A-(\partial_{\mu}{\bar A})A]\,. \label{v}
\end{eqnarray}
The Skyrme Lagrangian~(\ref{skyrme_action}) is then written as 
\begin{eqnarray}
	{\cal L}=-\frac{f_{\pi}^{2}}{8}{\bar D}^{\mu}{\bar A}D_{\mu}A
	+\frac{1}{16e^{2}}(B^{*}_{[\mu}B_{\nu]})^{2} \label{gauged_skyrme}
\end{eqnarray}
where  
\begin{eqnarray}
	B_{\mu}=i\epsilon^{ij}A_{i}\partial_{\mu}A_{j}\,. \label{}
\end{eqnarray}
Now, $\omega = U^{\dagger}\partial_{\mu}Udx^{\mu}$ is an SU(2)-valued one-form and therefore 
the Maurer-Cartan equation holds 
\begin{eqnarray}
	d\omega +\omega \wedge \omega =0 \,,\label{}
\end{eqnarray}
which reads 
\begin{eqnarray}
	F_{\mu\nu}\equiv \partial_{[\mu}V_{\nu]}=-iB^{*}_{[\mu}B_{\nu]}\,. \label{FB}
\end{eqnarray}
Thus, the Lagrangian~(\ref{gauged_skyrme}) becomes
\begin{eqnarray}
	{\cal L}=-\frac{f_{\pi}^{2}}{8}{\bar D}^{\mu}{\bar A}D_{\mu}A
	-\frac{1}{16e^{2}}F_{\mu\nu}^{2}\,. \label{gauged_skyrme1}
\end{eqnarray}

To supersymmetrise the model, let us extend $(A_{i}, V_{\mu})$ to the chiral multiplet 
and vector multiplet respectively  
\begin{eqnarray}
	A_{i}\rightarrow (A_{i}, \psi_{\alpha i}, F_{i})\;, 
	\;\;\; V_{\mu}\rightarrow (V_{\mu}, \lambda_{\alpha}, D)  \label{}
\end{eqnarray}
where $i, \alpha=1, 2$ and $F_{i}$ are complex scalars, $D$ is a real scalar 
and $\psi_{\alpha i}$, $\lambda_{\alpha}$ are Majorana spinors.
Then, the supersymmetric Lagrangian is given by 
\begin{eqnarray}
	{\cal L}_{SUSY}&=&\frac{f_{\pi}^{2}}{8}\left[-{\bar D}^{\mu}{\bar A}D_{\mu}A
	+\frac{i}{2}D_{\mu}\psi^{\alpha}\sigma^{\mu}_{\alpha{\dot \alpha}}
	{\bar \psi}^{{\dot \alpha}}-\frac{i}{2}\psi^{\alpha}
	\sigma_{\mu\alpha{\dot \alpha}}{\bar D}^{\mu}{\bar \psi}^{{\dot \alpha}} 
	+ {\bar F}F-i{\bar A}\lambda^{\alpha}\psi_{\alpha}
	+iA{\bar \lambda}_{{\dot \alpha}}{\bar \psi}^{{\dot \alpha}}
	+D({\bar A}A-1) \right] \nonumber \\
	&&+\frac{1}{8e^{2}}\left[-\frac{1}{2}F_{\mu\nu}^{2}
	-i\partial_{\mu}\lambda^{\alpha}\sigma^{\mu}_{\alpha{\dot \alpha}}
	{\bar \lambda}^{{\dot \alpha}}+D^{2}\right]\,. \label{susy_lag}
\end{eqnarray}
One can show that these are invariant under the following supersymmetric transformations: 
\begin{eqnarray}
	&\delta A_{i}&=-\epsilon^{\alpha}\psi_{\alpha i}\\
	&\delta \psi_{\alpha i}&=-i\sigma^{\mu}_{\alpha{\dot \alpha}}
	{\bar \epsilon}^{{\dot \alpha}}D_{\mu}A_{i}+\epsilon_{\alpha}F_{i}\\
	&\delta F_{i}&=-i{\bar \epsilon}^{{\dot \alpha}}
	{\bar \sigma}^{\mu}_{{\dot \alpha}\alpha}D_{\mu}\psi^{\alpha}_{i}
	-i{\bar \epsilon}_{{\dot \alpha}}A_{i}
	{\bar \lambda}^{{\dot \alpha}} \label{}
\end{eqnarray}
and 
\begin{eqnarray}
	&\delta V_{\mu}&=-\frac{i}{2}({\bar \epsilon}_{{\dot \alpha}}
	{\bar \sigma}_{\mu}^{{\dot \alpha}\alpha}\lambda_{\alpha}
	+\epsilon_{\alpha}\sigma_{\mu}^{\alpha{\dot \alpha}}
	{\bar \lambda}_{{\dot \alpha}}) \\
	&\delta \lambda_{\alpha}&=-\epsilon^{\beta}\sigma^{\mu\nu}_{\beta\alpha}F_{\mu\nu}
	+i\epsilon_{\alpha}D \\
	&\delta D&=\frac{1}{2}({\bar \epsilon}^{{\dot \alpha}}
	{\bar \sigma}^{\mu}_{{\dot \alpha}\alpha}\partial_{\mu}\lambda^{\alpha}
	-\epsilon^{\alpha}\sigma^{\mu}_{\alpha{\dot \alpha}}
	\partial_{\mu}{\bar \lambda}^{{\dot \alpha}})\,.\label{}
\end{eqnarray}
With these transformations and the field equations, the constraints ${\bar A}A=1$ can be 
extended to    
\begin{eqnarray}
	{\bar A}A=1\;, \;\;\; {\bar A}\psi_{\alpha}=0 \;,\;\;\; {\bar A}F=0\,, \label{}
\end{eqnarray}
and resultantly one obtains
\begin{eqnarray}
	&V_{\mu}&=-\frac{i}{2}({\bar A}\partial_{\mu}A-\partial_{\mu}{\bar A}A)
	+\frac{1}{2}\psi^{\alpha} \sigma_{\mu\alpha{\dot \alpha}}{\bar \psi}^{{\dot \alpha}} \\
	&\lambda_{\alpha}&=-i{\bar F}\psi_{\alpha}+\sigma^{\mu}_{\alpha {\dot \alpha}}
	{\bar \psi}^{{\dot \alpha}}D_{\mu}A \\
	&D&={\bar D}^{\mu}{\bar A}D_{\mu}A-\frac{i}{2}(D_{\mu}\psi^{\alpha}
	\sigma^{\mu}_{\alpha{\dot \alpha}}{\bar \psi}^{{\dot \alpha}}-\psi^{\alpha}
	\sigma_{\mu\alpha{\dot \alpha}}{\bar D}^{\mu}{\bar \psi}^{{\dot \alpha}})\,. \label{}
\end{eqnarray}
Setting $\psi_{\alpha}=F_{i}=0$, one arrives at the bosonic sector of Eq.~(\ref{susy_lag}), 
\begin{eqnarray}
	{\cal L}_{SUSY}=-\frac{f_{\pi}^{2}}{8}{\bar D}^{\mu}{\bar A}D_{\mu}A
	+\frac{1}{8e^{2}}\left[-\frac{1}{2}F_{\mu\nu}^{2}
	+({\bar D}^{\mu}{\bar A}D_{\mu}A)^{2}\right]\,. \label{}
\end{eqnarray}
The last term is fourth-order in derivatives. There are other possible 
fourth-order terms we can add to the Lagrangian, which turn out~\cite{bergshoeff85}  
\begin{eqnarray}
	\square {\bar A}\square A -({\bar D}^{\mu}{\bar A}D_{\mu}A)^{2}\,. \label{}
\end{eqnarray}
The most general supersymmetric Skyrme model is thus given by 
\begin{eqnarray}
	{\cal L}_{SUSY}=-\frac{f_{\pi}^{2}}{8}{\bar D}^{\mu}{\bar A}D_{\mu}A
	+\frac{1}{8e^{2}}\left[\alpha \left\{-\frac{1}{2}F_{\mu\nu}^{2}
	+({\bar D}^{\mu}{\bar A}D_{\mu}A)^{2}\right\} 
	+\beta\left\{\frac{}{}\square {\bar A}\square A -({\bar D}^{\mu}{\bar A}
	D_{\mu}A)^{2}\right\} \right]\,. \label{susy_lag1}
\end{eqnarray}
Although we have simply set $F_{i}=0$ to get Eq.~(\ref{susy_lag1}), 
the Lagrangian~(\ref{susy_lag}) contains derivatives of $F_{i}$ 
and hence $F_{i}$ becomes a dynamical field. 
However, since the dynamical solution does not cancel the higher derivative 
terms to recover Eq.~(\ref{gauged_skyrme1}), the Lagrangian would be still 
given in the form of~(\ref{susy_lag1}).  

We consider spherically symmetric solutions with the topological charge $1$. 
Let us impose the hedgehog ansatz 
\begin{eqnarray}
	U=\cos f(r) + i{\vec \tau}\cdot {\vec n}\sin f(r) \label{}
\end{eqnarray}
where ${\vec n}={\vec x}/r$. In terms of $A$, it corresponds to 
\begin{eqnarray}
	A_{1}=\cos f(r) +i\cos\theta\sin f(r) \;, \;\;\;
	A_{2}=ie^{i\varphi}\sin\theta\sin f(r) \,. \label{ansatz}
\end{eqnarray}
The topological charge in the Skyrme model is defined by 
\begin{eqnarray}
	Q=-\frac{1}{24\pi^{2}}\int d^{3}x \, \epsilon^{ijk}\,{\rm tr}
	(U^{\dagger}\partial_{i}UU^{\dagger}\partial_{j}U
	U^{\dagger}\partial_{k}U)=-\frac{1}{8\pi^{2}}\int d^{3}x \, 
	\epsilon^{ijk}V_{i}F_{jk}\,. \label{q}
\end{eqnarray}
where we have used the relation~(\ref{FB}) in the equality. 
Inserting the ansatz~(\ref{ansatz}) into (\ref{q}) and using the boundary conditions $f(0)=\pi$, 
$f(\infty)=0$, one obtains $Q=1$.

We insert the ansatz~(\ref{ansatz}) into the Lagrangian~(\ref{susy_lag1}). 
Then the static energy is given by
\begin{eqnarray}
	E&=&4\pi\frac{f_{\pi}}{e}\int dx\,x^{2}\left[\frac{1}{12}\left\{(f')^{2}
	+\frac{2\sin^{2}f}{x^{2}}\right\}+\frac{(\alpha + \beta)}{15}\left\{(f')^{2}
	-\frac{\sin^{2}f}{x^{2}}\right\}^{2}+\frac{\beta}{12}\left(f''+\frac{2f'}{x}
	-\frac{\sin 2f}{x^{2}}\right)^{2}\right]\,. \label{}
\end{eqnarray}
where we have introduced the dimensionless variable $x=ef_{\pi}r$ and the prime denotes 
a derivative with respect to $x$. 
 
\section{3. Field Equations}   
 
The field equation can be obtained by taking a variation with respect to $f(x)$,  
\begin{eqnarray}
	&& -x^{2}f''-2xf'+\sin 2f+\frac{4(\alpha+\beta)}{5}\left[2f''\sin^{2}f
	-6x^{2}(f')^{2}f''-4x(f')^{3}+(f')^{2}\sin 2f+\frac{\sin^{2}f\sin 2f}{x^{2}}
	\right] \nonumber \\
	&& +\beta\left[x^{2}f^{(4)}+4xf^{(3)}-4f''\cos 2f+4(f')^{2}\sin 2f
	-\frac{4\sin^{2}f\sin2f}{x^{2}}\right]=0 \,. \label{field-eq}
\end{eqnarray}
This equation contains third- and fourth-order derivative terms. 
In the theories which contain higher-derivative terms, there 
exists no lowest-energy state no matter how small their coefficients are. 
Thus, if one solves them directly, one would end up with picking up unphysical 
runaway solutions. The perturbative method to avoid this problem was proposed 
in Ref.~\cite{simon90} and has been applied for theories such as gravity and 
knotted solitons~\cite{sawado06}.  

Let us apply the perturbation to solve the equation~(\ref{field-eq}). 
The Skyrme model is a truncated theory of derivative expansion. 
Therefore, the higher-derivative terms would be treated as small corrections. 
Let us expand the Skyrme field in the order of the coefficient of the higher-derivative 
terms $\beta$, 
\begin{eqnarray}
	f(x)=f_{0}(x)+\sum_{l=1}^{L}\beta^{l} f_{l}(x)+O(\beta^{L+1})\,. \label{}
\end{eqnarray}
We compute solutions upto the second-order in $\beta$, {\it i.e.} 
$L=2$. 

From Eq.~(\ref{field-eq}), the zeroth-order equation is given by 
\begin{eqnarray}
	O(\beta^{0}) \;\; : \;\; 
	h_{0}f_{0}''-2rf_{0}'+\sin 2f_{0}+\frac{4\alpha}{5}\left[-4r(f_{0}')^{3}
	+(f_{0}')^{2}\sin 2f_{0}+\frac{\sin^{2}f_{0}\sin 2f_{0}}{r^{2}}\right]=0 \label{f0}
\end{eqnarray}
where 
\begin{eqnarray}
	h_{0}=-r^{2}+\frac{8\alpha}{5}\left(\sin^{2}f_{0}-3r^{2}f_{0}'^{2}\right)\,. \label{}
\end{eqnarray}
The first-order equation is given by 
\begin{eqnarray}
	O(\beta) \;\; : \;\; 
	h_{0}f_{1}''-s_{1}f_{1}'+s_{2}f_{1}+s_{3}+\frac{4}{5}s_{4}=0 \label{f1}
\end{eqnarray}
where 
\begin{eqnarray}
	s_{1}&=&2r+\frac{8\alpha}{5}\left(\frac{}{}6r^{2}f_{0}'f_{0}''
	+6rf_{0}'^{2}-f_{0}'\sin 2f_{0}\right)\\
	s_{2}&=&2\cos 2f_{0}+\frac{4\alpha}{5}\left(\frac{}{}
	2f_{0}''\sin 2f_{0}+2f_{0}'^{2}\cos 2f_{0} +w_{0}\right)\\
	s_{3}&=&r^{2}f_{0}^{(4)}+4rf_{0}^{(3)}-4f_{0}''\cos 2f_{0}
	+4f_{0}'^{2}\sin 2f_{0}-\frac{4\sin^{2}f_{0}\sin 2f_{0}}{r^{2}}\\
	s_{4}&=&2(\sin^{2}f_{0}-3r^{2}f_{0}'^{2})f_{0}''
	-4rf_{0}'^{3}+f_{0}'^{2}\sin 2f_{0}+\frac{\sin^{2}f_{0}\sin 2f_{0}}{r^{2}}
\end{eqnarray}
and 
\begin{eqnarray}
	w_{0}=\frac{1}{r^{2}}(\sin^{2}2f_{0}+2\sin^{2}f_{0}\cos 2f_{0})\,.
\end{eqnarray}
The second-order equation is given by 
\begin{eqnarray}
	O(\beta^{2}) \;\; : \;\;
	h_{0}f_{2}''+\frac{4\alpha}{5}(2m_{1}+m_{2})+m_{3}+\frac{4}{5}m_{4}+m_{5}=0 \label{f2}
\end{eqnarray}
where 
\begin{eqnarray}
	m_{1}&=&-6r^{2}f_{0}'f_{0}''f_{2}'+f_{0}''\sin 2f_{0}f_{2}+(\sin 2f_{0}f_{1}
	-6r^{2}f_{0}'f_{1}')f_{1}''+f_{1}^{2}f_{0}''\cos 2f_{0}-3r^{2}f_{1}'^{2}f_{0}''\\
	m_{2}&=&2(f_{0}'\sin 2f_{0}-6rf_{0}'^{2})f_{2}'+(2f_{0}'^{2}\cos 2f_{0}+w_{0})f_{2}
	+(\sin 2f_{0}-12rf_{0}')f_{1}'^{2}+4f_{0}'f_{1}'f_{1}\cos 2f_{0}\nonumber\\
	&&+\left[-2f_{0}'^{2}\sin 2f_{0}+\frac{1}{r^{2}}(3\sin 2f_{0}\cos 2f_{0}
	-2\sin^{2}f_{0}\sin 2f_{0})\right]f_{1}^{2}\\
	m_{3}&=&r^{2}f_{1}^{(4)}+4rf_{1}^{(3)}-4f_{1}''\cos 2f_{0}+8f_{0}'f_{1}'
	\sin 2f_{0}+4(2f_{0}''\sin 2f_{0}+2f_{0}'^{2}\cos 2f_{0}-w_{0})f_{1} \\
	m_{4}&=&2(\sin^{2}f_{0}-3r^{2}f_{0}'^{2})f_{1}''+2(f_{0}'\sin 2f_{0}
	-6rf_{0}'^{2}-6r^{2}f_{0}'f_{0}'')f_{1}'
	+(2f_{0}''\sin 2f_{0}+2f_{0}'^{2}\cos 2f_{0}+w_{0})f_{1} \\
	m_{5}&=&-2rf_{2}'+2f_{2}\cos 2f_{0}-2f_{1}^{2}\sin 2f_{0}\,.
\end{eqnarray}
Solutions upto the second-order are then given by 
\begin{eqnarray}
	f(r)=f_{0}(x)+\beta f_{1}(x)+\beta^{2}f_{2}(x)+O(\beta^{3})\,. \label{}
\end{eqnarray}

\section{4. Numerical Solutions}
\begin{figure}
\includegraphics[height=6.5cm, width=8.5cm]{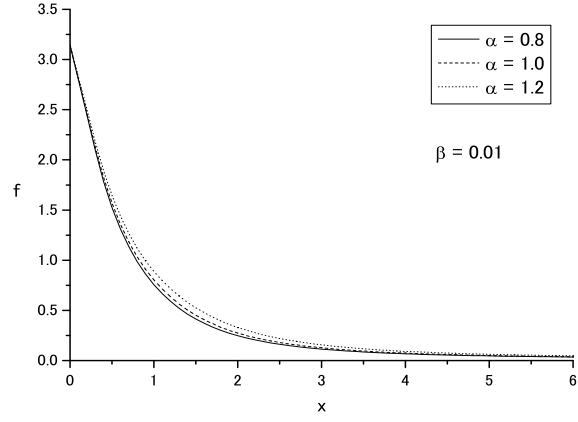}
\caption{\label{fig:fa} The profile function $f$ as a 
function of $x$ for with $\alpha =0.8, 1.0, 1.2$ and $\beta=0.01$ fixed.}
\end{figure}

\begin{figure}
\includegraphics[height=6.5cm, width=8.5cm]{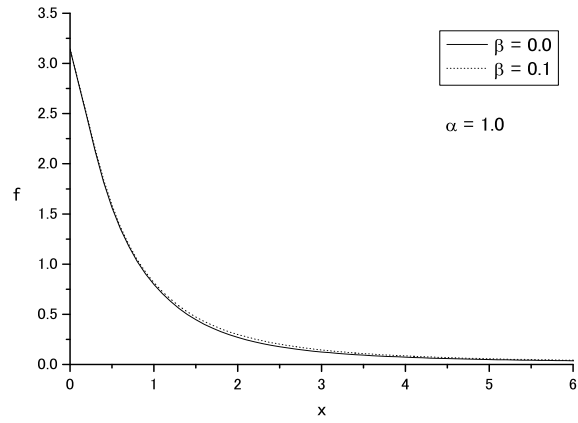}
\caption{\label{fig:fb} The profile function $f$ as a 
function of $r$ for $\beta = 0.0, 0.1$ with $\alpha =1.0$ fixed.}
\end{figure}

\begin{figure}
\includegraphics[height=6.5cm, width=8.5cm]{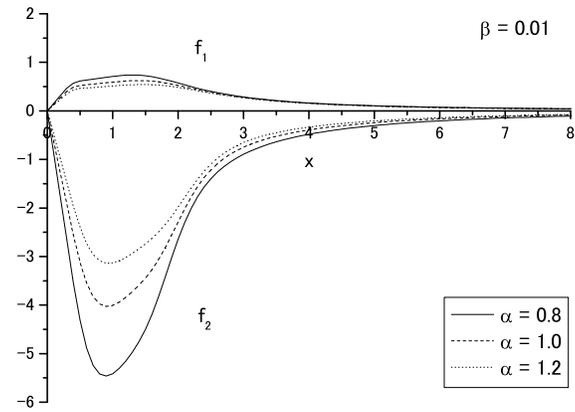}
\caption{\label{fig:f1f2} The perturbed profile function $f_{1}$ and 
$f_{2}$ as a function of $x$ for $\alpha =0.8, 1.0, 1.2$.}
\end{figure}

\begin{figure}
\includegraphics[height=6.5cm, width=8.5cm]{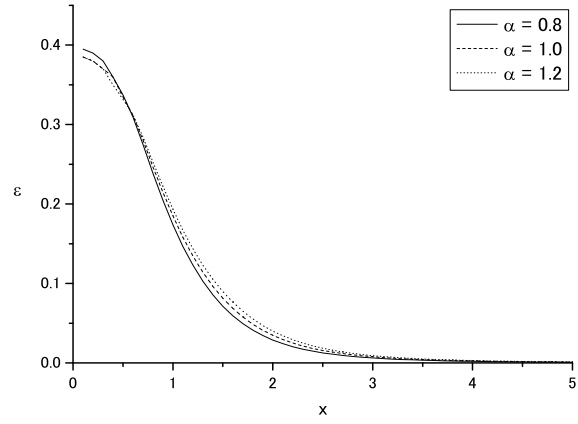}
\caption{\label{fig:ed} The energy density as a function of $x$ 
for $\alpha =0.8, 1.0, 1.2$.}
\end{figure}

We have solved the second-order differential equations~(\ref{f0},\ref{f1},\ref{f2}) 
by the shooting method subject to the boundary conditions 
\begin{eqnarray}
	f_{0}(0)=\pi \;, \;\;\; f_{0}(\infty)=0 \;, \;\;\; 
	f_{1}(0)=0 \;, \;\;\; f_{1}(\infty)=0 \;, \;\;\;
	f_{2}(0)=0 \;, \;\;\; f_{2}(\infty)=0 \,. \label{}
\end{eqnarray}
Fig.~\ref{fig:fa} shows the dependence of the profile function $f(x)$ on $\alpha$. 
The skyrmion slightly expands in size for increasing $\alpha$. 
Fig.~\ref{fig:fb} shows its dependence on $\beta$. The difference is very small but 
non-zero $\beta$ contributes to expand the size of skyrmions. 
Fig.~\ref{fig:f1f2} shows the dependence on $\alpha$ of the first- 
and second-order perturbed functions. The values which give a maximum correction 
to the leading order are $f_{1}\approx 1$ and $f_{2}\approx -5$ for $\alpha=0.8$. 
If we take $\beta=0.01$, each will give a correction 
$O(10^{-2})$ and $O(10^{-4})$ respectively. The leading order takes the value of 
$O(1)$ around the maximum correction values. Thus, it seems that $\beta=0.01$ 
would be small enough to make our perturbation method valid at least 
for $\alpha \gtrsim 0.8$. 
Fig.~\ref{fig:ed} shows the $\alpha$ dependence of the energy density. 
The behavior is the same as the profile, that is, the skyrmion increses 
in size as $\alpha$ increases. 

The corrections gets larger as $\alpha$ becomes smaller, and correspondingly 
we have to take smaller values for $\beta$. But the regime where $f_{1}$ and $f_{2}$ 
take large values compared to the leading order, the derivative expansion 
would be broken down and the result should not be trusted.  

It is noted that the first-order function contributes to increase 
the size of the skyrmion. On the other hand, the second-order 
function contributes to decrease the size. 
That makes the total corrections to the leading-order solution even smaller. 

\begin{figure}
\includegraphics[height=6.5cm, width=8.5cm]{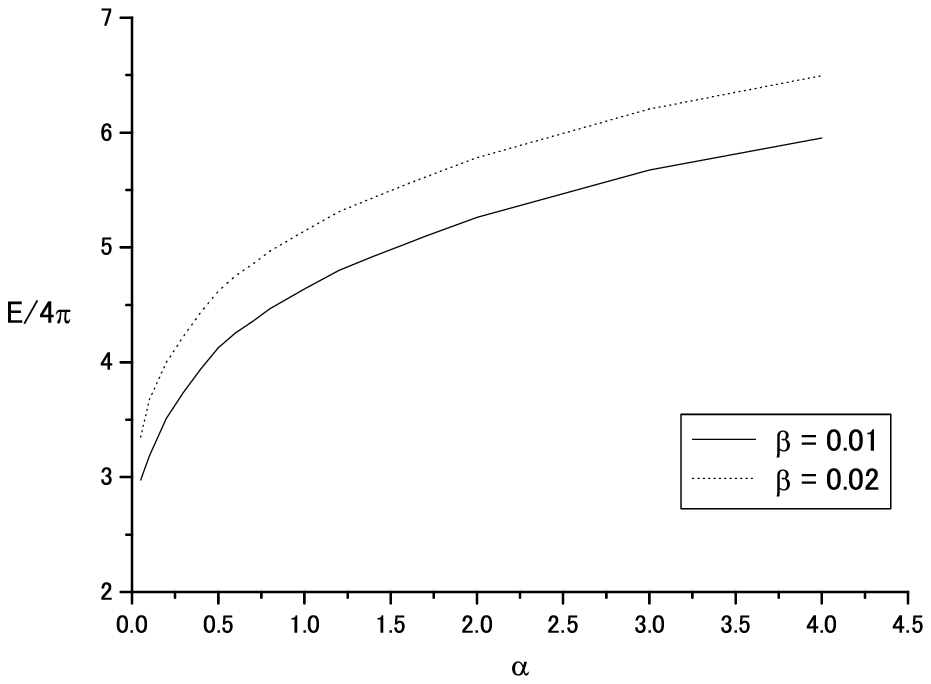}
\caption{\label{fig:energy} The $\alpha$ dependence of the 
total energy with $\beta=0.01$ fixed.}
\end{figure}

We have examined the dependence of the total energy on $\alpha$ and $\beta$ 
which is shown in Fig.~\ref{fig:energy}. 
For increasing $\alpha$, the energy increases monotonously. 
For increasing $\beta$, the energy increases and the figure  
merely shifts upwords.  

In Ref.~\cite{bergshoeff85}, it was shown that skyrmions are unstable when $\beta=0$.   
However, for $\beta \neq 0$, there is a possibility that they are stable. 
Unfortunately we are not able to show if our solutions are stable since 
the equation for the stability analysis are again higher-order 
in derivatives.

\section{5. Summary}
In this paper we studied the ${\cal N}=1$ supersymmetric Skyrme model and 
constructed skyrmion solutions numerically by the perturbation method upto 
the second order in the coefficient of the higher derivative terms. 
The skyrmion depends on the coupling constants $\alpha$ and $\beta$ 
and increases in size when these values increase. 
We found that the first-order correction contributes to increasing the size 
of the skyrmion while the second-order contributes to decreasing the size. 
As a result, the total correction to the leading-order solution is very small. 
For $\beta=0.01$, the first- and second-order gives a correction of $1\%$ and 
$0.01\%$ respectively to the leading order solution, which should justify 
our perturbative treatment of the model. 
The energy of the skyrmion increases monotonously for increasing 
$\alpha$ and $\beta$. 

It should be also noted that for $\beta=0$, the supersymmetric skyrmions are 
unstable, but the higher-derivative terms could change the stability dramatically 
no matter how small the correction is although we have to wait for the results 
from proper stability analysis,. 

We believe that the supersymmetric Skyrme model and its soliton solutions deserves 
further investigation since it may shed light on the non-perturbative 
effects in large-$N$ supersymmetric QCD theory. 
In particular, quntisation of supersymmetric skyrmions will be necessary 
to be performed.

\section{Acknowledgement}
We would like to thank Muneto Nitta for useful discussions and comments.  



\begin{thebibliography}{99}

\bibitem{skyrme58}
T. H. R. Skyrme, Proc. Roy. Soc. Lond. A247 (1958) 260. 

\bibitem{witten79}
E. Witten, Nucl. Phys. B160 (1979) 57. 

\bibitem{witten83}
E. Witten, Nucl. Phys. B223 (1983) 422; Nucl. Phys. B223 (1983) 433.

\bibitem{adkins83}
G. Adkins, C. Nappi and E. Witten, Nucl. Phys. B228 (1983) 552.

\bibitem{dvali99}
G. Dvali, G. Gabadadze and Z. Kakushadze, Nucl. Phys. B562 (1999) 158. 

\bibitem{witten98}
E. Witten, Phys. Rev. Lett. 81 (1998) 2862. 

\bibitem{shifman99}
M. Shifman, Phys. Rev. D59 (1999) 021501.

\bibitem{witten97}
E. Witten, Nucl. Phys. B507 (1997) 658.

\bibitem{gabadadze00}
G. Gabadadze and M. Shifman, Phys. Rev. D61 (2000) 075014; 

\bibitem{bergshoeff85}
E. A. Bergshoeff, R. I. Nepomechie, H. J. Schnitzer, Nucl. Phys. B249 (1985) 93. 

\bibitem{simon90}
J. Simon, Phys. Rev. D41 (1990) 3720. 

\bibitem{zumino79}
B. Zumino, Phys. Lett. B87 (1979) 203.

\bibitem{sawado06}
N. Sawado, N. Shiiki and S. Tanaka, SIGMA Vol.2 (2006) 016; 
hep-ph/0511208; hep-ph/0507258. 

\end{thebibliography}
\end{document}